\documentclass[a4paper,11pt]{article}
\usepackage{pos}

\title{Sparse modeling study to extract spectral functions from lattice QCD data}

\author*[a]{Junichi Takahashi}
\author[b]{Hiroshi Ohno}
\author[c]{Akio Tomiya}

\affiliation[a]{Meteorological College, Japan Meteorological Agency,\\
7-4-81, Asahi-cho, Kashiwa, Chiba 277-0852, Japan}
\affiliation[b]{Center for Computational Sciences, University of Tsukuba,\\
1-1-1, Tennodai, Tsukuba, Ibaraki 305-8577, Japan}
\affiliation[c]{Department of Information and Mathematical Sciences,
Tokyo Woman’s Christian University, \\
2-6-1, Zempukuji, Suginami-ku, Tokyo 167-8585, Japan}

\emailAdd{mhjkk-takahashi@met.kishou.go.jp}
\emailAdd{hohno@ccs.tsukuba.ac.jp}
\emailAdd{akio@yukawa.kyoto-u.ac.jp}

\abstract{
We present spectral functions extracted from Euclidean-time correlation functions
by using sparse modeling.
Sparse modeling is a method that solves inverse problems
by considering only the sparseness of the solution we seek.
To check applicability of the method,
we firstly test it with mock data
which imitate charmonium correlation functions on a fine lattice.
We show that the method can reconstruct the resonance peaks in the spectral functions.
Then, we extract charmonium spectral functions
from correlation functions obtained from lattice QCD
at temperatures below and above the critical temperature $T_{\mathrm{c}}$.
We show that this method yields results like those obtained with MEM and other methods.
}

\FullConference{The 41st International Symposium on Lattice Field Theory (LATTICE2024)\\
 28 July - 3 August 2024\\
Liverpool, UK\\}

\begin{document}
\maketitle

\section{Introduction}
Theoretically accessible real-frequency spectral functions are crucial
for the study of properties of the hot and dense medium such as the Quark-Gluon Plasma
~\cite{McLerran:1984ay,Matsui:1986dk,Braaten1990.PhysRevLett.64.2242,Petreczky2006.PhysRevD.73.014508}.
However,
the extraction of these spectral functions presents a significant challenge
due to the inability to obtain them directly from lattice QCD calculations.
\\
\indent
In lattice QCD,
an imaginary time description is employed
to calculate the correlation function $G$ of Euclidean time $\tau$,
which relates the spectral function $\rho$ of frequency $\omega$
through the following integral equation:
\begin{equation}
  G(\tau)
  =\int^{\infty}_{0}d\omega K(\omega,\tau)\rho(\omega),
  \label{eq:G=intKrho}
\end{equation}
where $K$ is the integration kernel defined by
\begin{equation}
  \displaystyle K(\omega,\tau)\equiv
  \frac{\cosh\left[
      \omega\left(
      \tau-\frac{1}{2T}
      \right)
      \right]}{\sinh\left(
    \frac{\omega}{2T}
    \right)}
  \label{eq:1_kernel}
\end{equation}
in the Euclidean time range $0\le\tau\le 1/T$ with temperature $T$.
Consequently,
in order to obtain spectral functions,
it is necessary to perform analytical continuations
from correlation functions.
In general,
the correlation functions obtained by lattice QCD contain noise,
and the analytical continuation is extremely sensitive to this noise.
\\
\indent
When the frequency $\omega$ is discretized,
eq.~\eqref{eq:G=intKrho} can be simply written as a linear equation
\begin{equation}
  \vec{G}=K\vec{\rho},
  \label{eq:linear_eq}
\end{equation}
where $\vec{G}$ and $\vec{\rho}$ are $M$ and $N$ dimensional vectors, respectively,
and $K$ is an $M \times N$ matrix.
For typical lattice QCD calculations the temporal lattice size, i.e.,
$M$ is of $O(10)$ while $N$ must be of $O(10^3)$ for sufficiently good resolution of the spectral function.
Therefore, solving eq.~\eqref{eq:linear_eq} to extract the spectral function
is an ill-posed inverse problem.
\\
\indent
There have been lots of previous studies on extracting spectral functions from lattice QCD data,
employing a variety of techniques based on different ideas
~\cite{ASAKAWA2001459,Ding2018.PhysRevD.97.094503,Brandt.PhysRevD.92.094510}.
Sparse modeling is one of such techniques, which was recently applied for the first time to lattice QCD data
to obtain spectral functions of the energy-momentum tensor and the shear viscosity~\cite{itou2020sparse}.
In Ref.~\cite{Takahashi:2023Fc},
we have considered the covariance
between different Euclidean times of the correlation function
when utilizing sparse modeling,
checked the applicability of the method
and extracted the charmonium spectral functions.
In this study we aim to conduct a more comprehensive investigation into the applicability of sparse modeling.
Furthermore, we also compare our results with those of one of the previous studies to properly estimate
the systematic uncertainty.

\section{Sparse modeling}
Extracting spectral functions by using sparse modeling
has been proposed in condensed matter physics
~\cite{Shinaoka.PhysRevB.96.035147,Otsuki.PhysRevE.95.061302}
(see also a review paper~\cite{doi:10.7566/JPSJ.89.012001} for detail).
The procedure of the sparse modeling is outlined in Ref.~\cite{Takahashi:2023Fc},
and we summarize the key characteristics of the sparse modeling
in the following three points.
\\
\indent
First,
the reduction in rank of the spectral functions and correlation functions
is achieved by the exclusion of the contribution of small singular values
after their bases are transformed as follows,
\begin{equation}
    \vec{G}^{\prime}\equiv U^{\mathrm{t}}\vec{G},
    \quad
    \vec{\rho}^{\mathrm{t}}\equiv V^{\mathrm{t}}\vec{\rho},
\end{equation}
where $U$ and $V$ are $M\times M$ and $N\times N$ orthogonal matrices,
respectively,
obtained through the singular value decomposition of the kernel $K$,
\begin{equation}
    K=USV^{\mathrm{t}},
\end{equation}
where $S$ is a diagonal matrix composed of singular values.
In this study,
we retain only the components of $\vec{\rho}^{\prime}$ and $\vec{G}^{\prime}$
that fulfill the condition $s_{l}/s_{1}\ge 10^{-15}$
where $s_{l}$ is the $l$-th largest singular value.
The number of components that satisfy this condition is denoted as $L$.
Thus the rank of the retained components is $L$
and the size of $U$, $V$ and $S$ become $M\times L$,
$N\times L$ and $L\times L$,
respectively.
\\
\indent
Second,
an $\mathrm{L}_{1}$ regularization term is incorporated into the cost function
based on the square error,
making the optimization problem
a form of Least Absolute Shrinkage and Selection Operator (LASSO).
The cost function that we seek to minimise can be expressed as follows,
\begin{equation}
    F(\vec{\rho}^{\prime})
    =\frac{1}{2}(\vec{G}^{\prime}-S\vec{\rho}^{\prime})^2
    +\lambda ||\vec{\rho}^{\prime}||_1,
    \label{eq:F}
\end{equation}
where $||\cdot||_{1}$ stands for the $\mathrm{L}_{1}$ norm
defined by $||\vec{\rho}^{\prime}||_{1}\equiv\sum^{L}_{i=1}|\rho^{\prime}_{i}|$
and $\lambda$ is a positive hyperparameter
which controls the contribution of the $\mathrm{L}_{1}$ regularization relative to the square error.
This regularization promotes sparsity in the solution of $\vec{\rho}^{\prime}$.
\\
\indent
Third, 
this optimization problem is solved iteratively 
by alternating direction method of multipliers (ADMM) algorithm~\cite{Boyd.MAL-016}.
In this study,
the problem is solved for multiple values of $\lambda$,
and the most likely spectral function $\vec{\rho}$ is determined at the optimal value $\lambda_{\mathrm{opt}}$.
The estimation of $\lambda_{\mathrm{opt}}$ follows the procedure used in previous work~\cite{itou2020sparse}.

\section{Mock data tests}
Before we apply the sparse modeling to the actual lattice QCD data,
we test it with mock data which imitate possible charmonium spectral functions.
\\
\indent
Following in Ref.~\cite{Ding2018.PhysRevD.97.094503},
we make the input spectral functions
at temperatures below and above $T_{\mathrm{c}}$:
\begin{itemize}
    \item For $T<T_{\mathrm{c}}$,\\
        $\hat{\rho}_{\mathrm{below}}(\hat{\omega})
        =\tilde{\Theta}(\hat{\omega},\hat{\omega}_{1},\Delta_{1})
        (1-\tilde{\Theta}(\hat{\omega},\hat{\omega}_{2},\Delta_{2}))
        \hat{\rho}_{\mathrm{res}}
        +\tilde{\Theta}(\hat{\omega},\hat{\omega}_{3},\Delta_{3})
        \hat{\rho}_{\mathrm{Wilson}}$.
    \item For $T>T_{\mathrm{c}}$,\\
        $\hat{\rho}_{\mathrm{above}}(\hat{\omega})
        =\hat{\rho}_{\mathrm{trans}}
        +\tilde{\Theta}(\hat{\omega},\hat{\omega}_{4},\Delta_{4})
        (1-\tilde{\Theta}(\hat{\omega},\hat{\omega}_{5},\Delta_{5}))
        \hat{\rho}_{\mathrm{res}}
        +\tilde{\Theta}(\hat{\omega},\hat{\omega}_{6},\Delta_{6})
        \hat{\rho}_{\mathrm{Wilson}}$.
\end{itemize}
Here $\hat{\rho}_{\mathrm{res}}$, $\hat{\rho}_{\mathrm{trans}}$
and $\hat{\rho}_{\mathrm{Wilson}}$
denote a resonance peak, a transport peak and a free Wilson spectral function,
respectively,
and the hatted letters are dimensionless quantities.
The function $\tilde{\Theta}$ is a modified $\Theta$ function
introduced to smoothly connect each spectral function.
In Ref.~\cite{Ding2018.PhysRevD.97.094503},
free continuum spectral function is employed
in $\hat{\rho}_{\mathrm{below}}$.
In this study,
however,
free Wilson spectral function is utilized to account
for lattice cutoff effects at temperature below $T_{\mathrm{c}}$.
For details on the functional forms of $\hat{\rho}_{\mathrm{res}}$,
$\hat{\rho}_{\mathrm{trans}}$, $\hat{\rho}_{\mathrm{Wilson}}$ and $\tilde{\Theta}$,
see Ref.~\cite{Ding2018.PhysRevD.97.094503}.

\indent
The values of the parameters used in the above spectral functions
are summarized in table~\ref{tab:params_mock_spf}.
The position of the resonance peak is about $J/\psi$ meson mass ($\sim 3.1$ GeV)
for $T < T_{\mathrm{c}}$,
and a transport peak appears and the resonance peak becomes broader
for $T > T_{\mathrm{c}}$.
The values of the parameters used in the modified $\Theta$ function
are same as Ref.~\cite{Ding2018.PhysRevD.97.094503}.
In this study,
we consider that the range of frequencies $\hat{\omega}$ is from 0 to 4
with 8,001 points in $\hat{\omega}$-space.
\begin{table*}[tbp]
    \centering
    \begin{tabular}{l|c}
    \hline
        Spectral function & Parameters \\
    \hline
    \hline
        $\hat{\rho}_{\mathrm{res}}$ for $\hat{\rho}_{\mathrm{below}}$ & $c_{\mathrm{res}}=0.08/7$, $\Gamma=0.05$, $M=0.155$ \\
        $\hat{\rho}_{\mathrm{Wilson}}$ for $\hat{\rho}_{\mathrm{below}}$ & $c_{\mathrm{Wilson}}=0.5$, $b^{(1)}=2$, $b^{(2)}=1$, $m=0.073$, $N_{c}=3$, $N_{\sigma}=4096$\\
        $\hat{\rho}_{\mathrm{trans}}$ for $\hat{\rho}_{\mathrm{above}}$ & $c_{\mathrm{trans}}=5\times 10^{-5}$, $\eta=0.006$\\
        $\hat{\rho}_{\mathrm{res}}$ for $\hat{\rho}_{\mathrm{above}}$ & $c_{\mathrm{res}}=0.06$, $\Gamma=0.15$, $M=0.225$ \\
        $\hat{\rho}_{\mathrm{Wilson}}$ for $\hat{\rho}_{\mathrm{above}}$ & $c_{\mathrm{Wilson}}=1$, $b^{(1)}=3$, $b^{(2)}=1$, $m=0.073$, $N_{c}=3$, $N_{\sigma}=4096$\\
    \hline
    \end{tabular}
    \caption{
    Parameters for the mock spectral functions.
    See Ref.~\cite{Ding2018.PhysRevD.97.094503} for notations.
    }
    \label{tab:params_mock_spf}
\end{table*}

\begin{figure}[tbp]
  \centering
  \begin{minipage}{.49\textwidth}
    \includegraphics[width=1.0\linewidth]{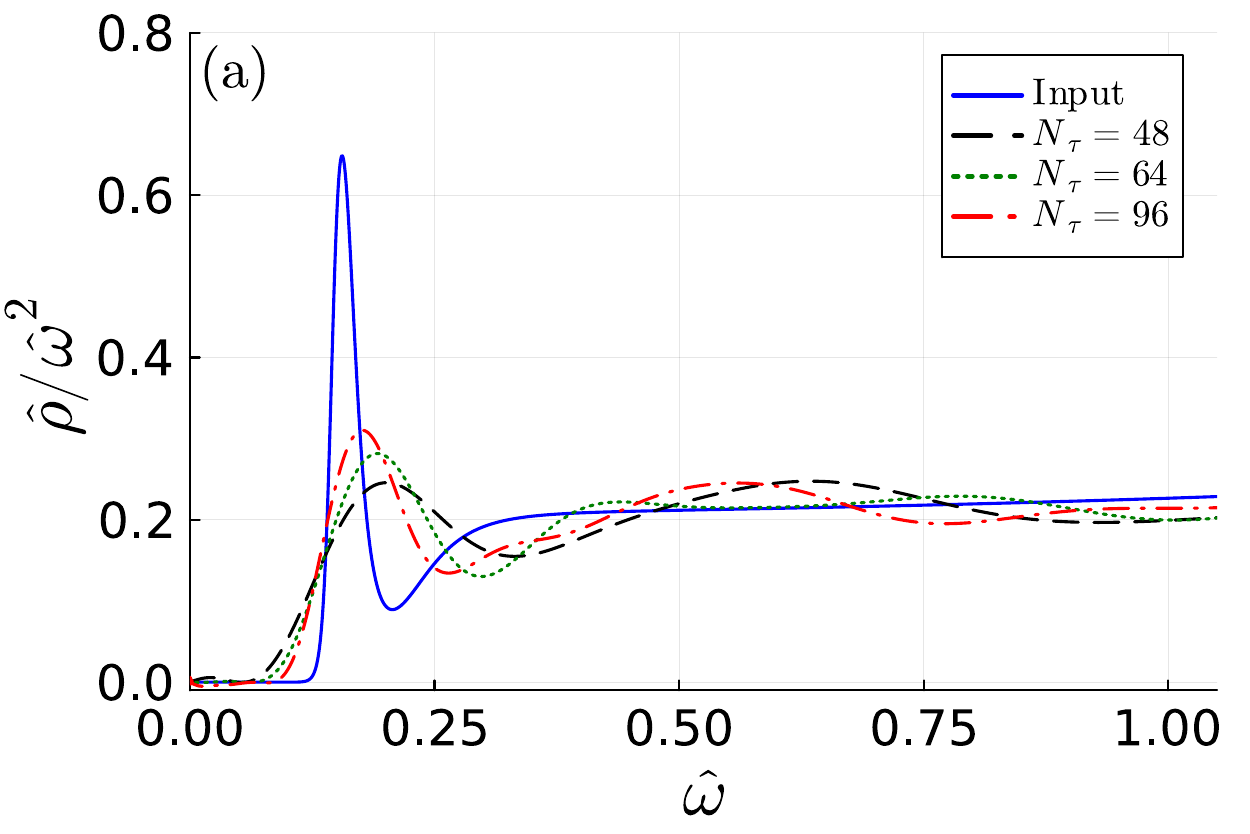}
  \end{minipage}
  \begin{minipage}{.49\textwidth}
    \includegraphics[width=1.0\linewidth]{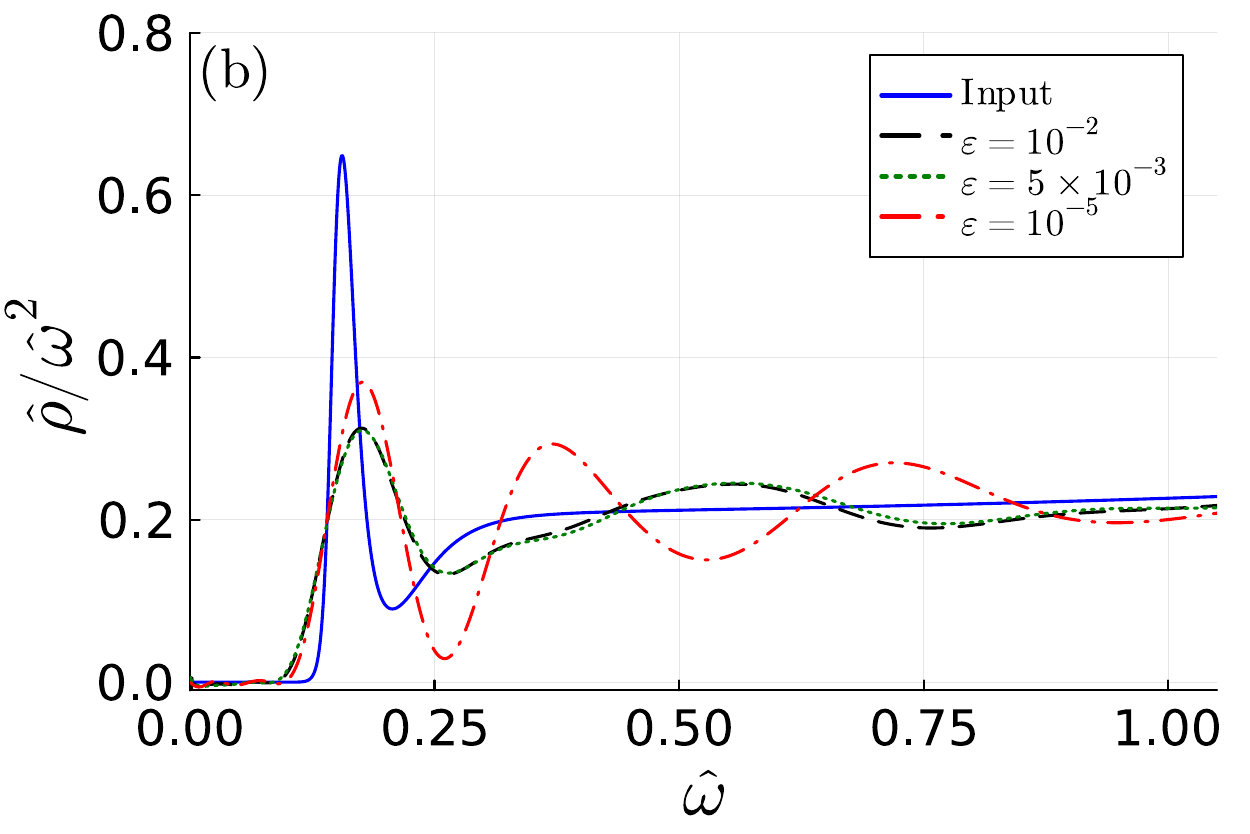}
  \end{minipage}
  \caption{
    Spectral functions calculated by using sparse modeling
    in the mock-data tests for $T<T_{\mathrm{c}}$.
    Figure (a) shows the results with a fixed noise level of $\varepsilon=5\times 10^{-3}$.
    The black dashed, green dotted and red dash-dotted lines 
    represent the output results with $N_{\tau}=48$, 64 and 96,
    respectively.
    Figure (b) shows the results with a fixed temporal extent of $N_{\tau}=96$.
    The black dashed, green dotted and red dash-dotted lines
    represent the output results with $\varepsilon=10^{-2}$, $5\times 10^{-3}$
    and $10^{-5}$,
    respectively.
    In both figures, 
    the blue solid line represents the input spectral function.
  }
  \label{fig:spf_mock_data_tests_bTc}
\end{figure}

\indent
The central values of correlation function $G(\tau)$ are given
by integrating the input spectral function and the kernel.
The kernel is given in eq.~\eqref{eq:1_kernel},
which diverges at $\omega=0$.
Moreover,
the correlation function is influenced by lattice cutoff effects at small $\tau$ distances.
To address these issues,
we used a modified kernel and a modified spectral function defined by
\begin{align}
  \tilde{K}(\hat{\omega},\hat{\tau};\hat{\tau}_{0})
  \equiv\displaystyle\hat{\omega}\frac{K(\hat{\omega},\hat{\tau})}{K(\hat{\omega},\hat{\tau}_{0})}
  =\hat{\omega}\frac{\cosh\left[
      \hat{\omega}\left(
      \hat{\tau}-\frac{N_{\tau}}{2}
      \right)
      \right]}{\cosh\left[
      \hat{\omega}\left(
      \hat{\tau}_{0}-\frac{N_{\tau}}{2}
      \right)
      \right]},
  \quad
  \tilde{\rho}(\hat{\omega};\hat{\tau}_{0})
  =\frac{\hat{\rho}(\hat{\omega})}{\hat{\omega}}K(\hat{\omega},\hat{\tau}_{0}),
\end{align}
and we used the mock correlation function data from $\hat{\tau}_{0}$ to $N_{\tau}/2$,
where $\hat{\tau}_{0}$ was set to 1 in our mock data tests.
Errors of $G(\tau)$ are generated
by gaussian random numbers with the variance
$\sigma(\tau)=\varepsilon\cdot\tau\cdot G(\tau)$
in order to incorporate the fact that
the error of lattice correlation functions increases as $\tau$ increases.
One of the purpose of this study
is to examine the applicability of sparse modeling
to the number of input data and the magnitude of noise.
Therefore,
we consider three types of temporal extents, $N_{\tau}=48,\;64$ and 96,
and three types of noise levels,
$\varepsilon=10^{-2},\;5\times 10^{-3}$ and $10^{-5}$
in our mock data tests.
\\
\indent
Figure~\ref{fig:spf_mock_data_tests_bTc} shows
the spectral functions as a function of $\hat{\omega}$
for $T<T_{\mathrm{c}}$.
In Fig.~\ref{fig:spf_mock_data_tests_bTc}(a),
the results with a fixed noise level of $\varepsilon=5\times 10^{-3}$
for various $N_{\tau}$ are illustrated.
The blue solid line represents the input spectral function,
and the black dashed, green dotted and red dash-dotted lines 
represent the output results with $N_{\tau}=48$, 64 and 96,
respectively.
Increasing $N_{\tau}$ results in better spectral functions
that are closer to the input spectral function.
Similar results are obtained for the other noise levels $\varepsilon$.
In Fig.~\ref{fig:spf_mock_data_tests_bTc}(b),
the results with a fixed temporal extent of $N_{\tau}=96$
for various $\varepsilon$ are illustrated.
The black dashed, green dotted and red dash-dotted lines
represent the output results
with $\varepsilon=10^{-2}$, $5\times 10^{-3}$ and $10^{-5}$,
respectively.
The meaning of the blue solid line is same as Fig.~\ref{fig:spf_mock_data_tests_bTc}(a).
Reducing $\varepsilon$ results in better spectral functions
that are closer to the input spectral function.
Similar results are obtained for the other temporal extents $N_{\tau}$.

\begin{figure}[tbp]
  \centering
  \begin{minipage}{.49\textwidth}
    \includegraphics[width=1.0\linewidth]{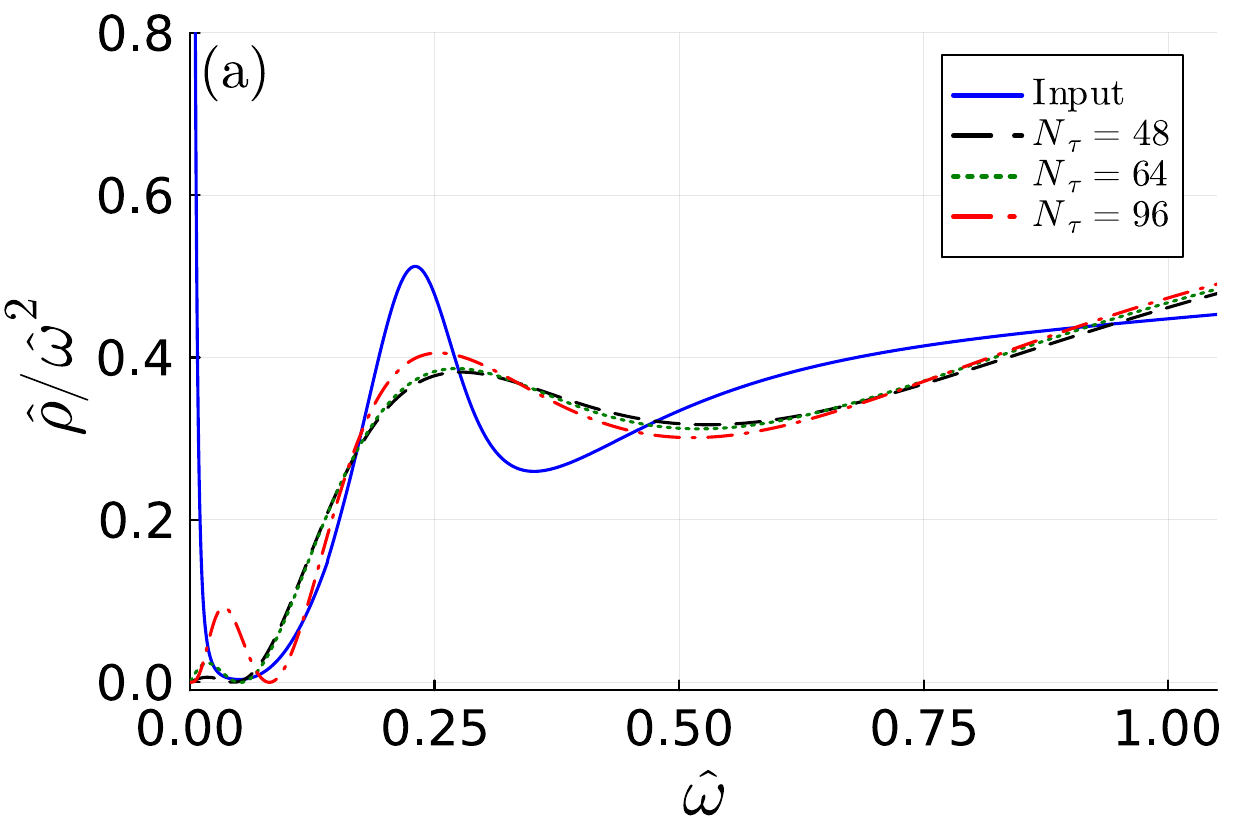}
  \end{minipage}
  \begin{minipage}{.49\textwidth}
    \includegraphics[width=1.0\linewidth]{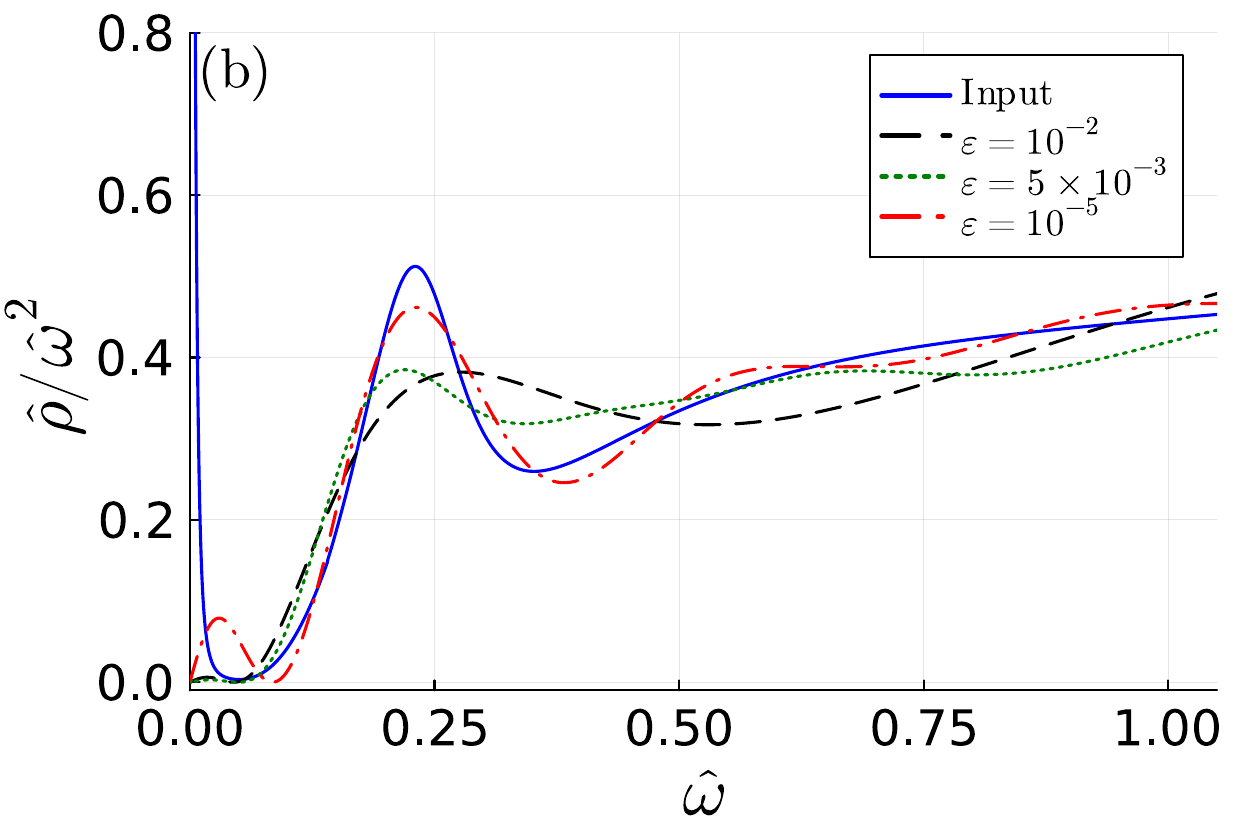}
  \end{minipage}
  \caption{
    Same as Fig.~\ref{fig:spf_mock_data_tests_bTc} but for $T>T_{\mathrm{c}}$.
    Figure (a) and (b) show the results
    with $\varepsilon=10^{-2}$ and $N_{\tau}=48$,
    respectively.
  }
  \label{fig:spf_mock_data_tests_aTc}
\end{figure}

\indent
Same results as Fig.~\ref{fig:spf_mock_data_tests_bTc}
but for $T<T_{\mathrm{c}}$ are shown in Fig.~\ref{fig:spf_mock_data_tests_aTc},
where $\varepsilon=10^{-2}$ and $N_{\tau}=48$ are chosen
in Fig.~\ref{fig:spf_mock_data_tests_aTc}(a)
and Fig.~\ref{fig:spf_mock_data_tests_aTc}(b),
respectively.
$\varepsilon$ and $N_{\tau}$ dependences are also the same
as Fig.~\ref{fig:spf_mock_data_tests_bTc},
i.e.,
larger $N_{\tau}$ and smaller $\varepsilon$ make the output closer to the input.
Nevertheless,
no transport peaks appear in the calculations
for any $N_{\tau}$ and $\varepsilon$ combinations.

\section{Results from lattice QCD data}
Next, we extracted the spectral function from actual lattice QCD data.
\\
\indent
We used the lattice data given in ref.~\cite{Ding2012.PhysRevD.86.014509},
where the correlation functions were measured with the $O(a)$-improved Wilson quark action
on quenched gauge configurations generated by using the standard plaquette gauge action.
The lattice spacing $a=0.010$ fm
and the corresponding $a^{-1}$ is about 18.97 GeV.
The spatial extent $N_{\sigma}$, 
the temporal extent $N_{\tau}$,
corresponding temperatures 
and the numbers of gauge configurations
are summarized in table~\ref{tab:4_LQCD_param}.
We utilized meson correlation functions
in the vector and the pseudoscalar channels
for each temperature.
We set $\hat{\tau}_{0}=4$,
and used the correlation function data from $\hat{\tau}_{0}$ to $N_{\tau}/2$.
\begin{table}[tpb]
  \centering
  \begin{tabular}{|c|c|c|c|}
    \hline
    $N_{\sigma}$ & $N_{\tau}$ & $T/T_{\mathrm{c}}$ & \# of conf.  \\
    \hline
    128          &  96        & 0.73               & 234          \\
    \hline
    128          &  48        & 1.46               & 461          \\
    \hline
  \end{tabular}
  \caption{
    The values of $N_{\sigma}$, $N_{\tau}$, corresponding temperatures,
    and number of configurations.
  }
  \label{tab:4_LQCD_param}
\end{table}

\begin{figure}[tbp]
  \centering
  \begin{minipage}{.49\textwidth}
    \includegraphics[width=1.0\linewidth]{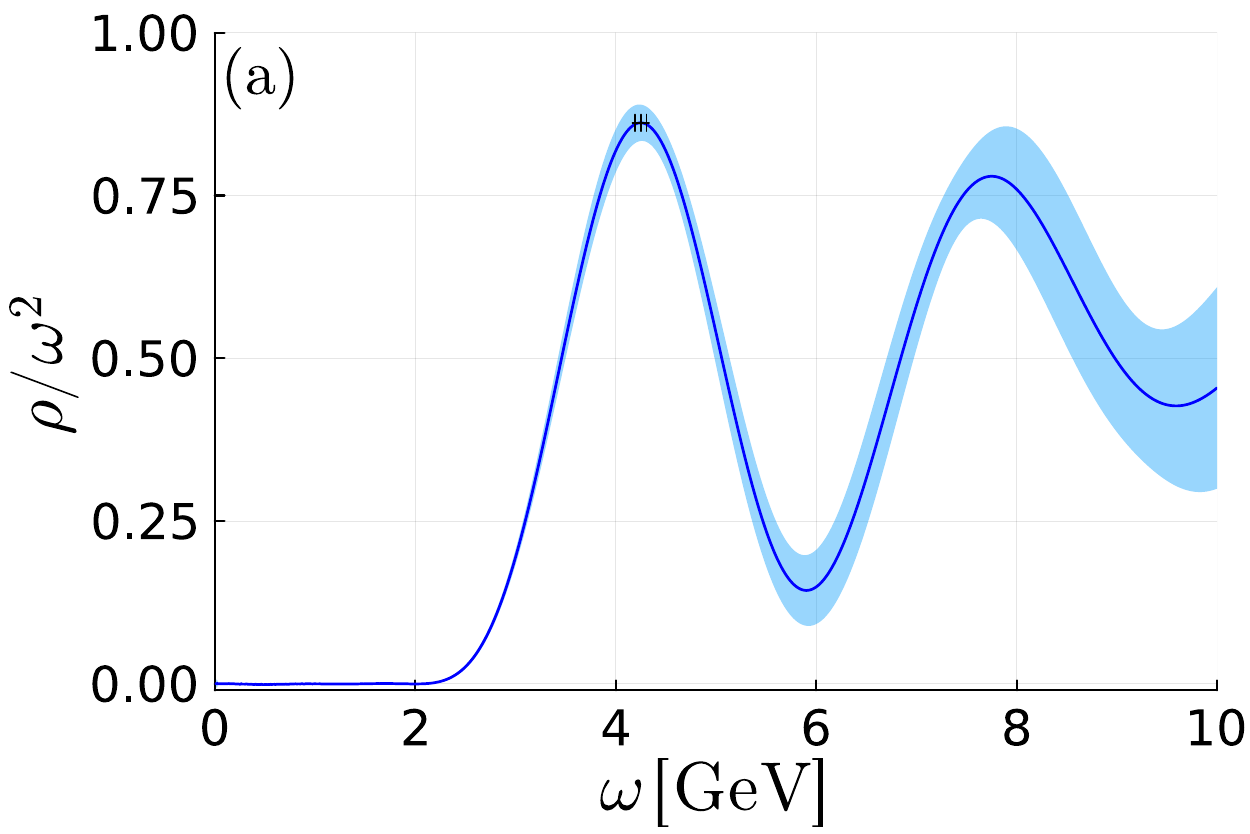}
  \end{minipage}
  \begin{minipage}{.49\textwidth}
    \includegraphics[width=1.0\linewidth]{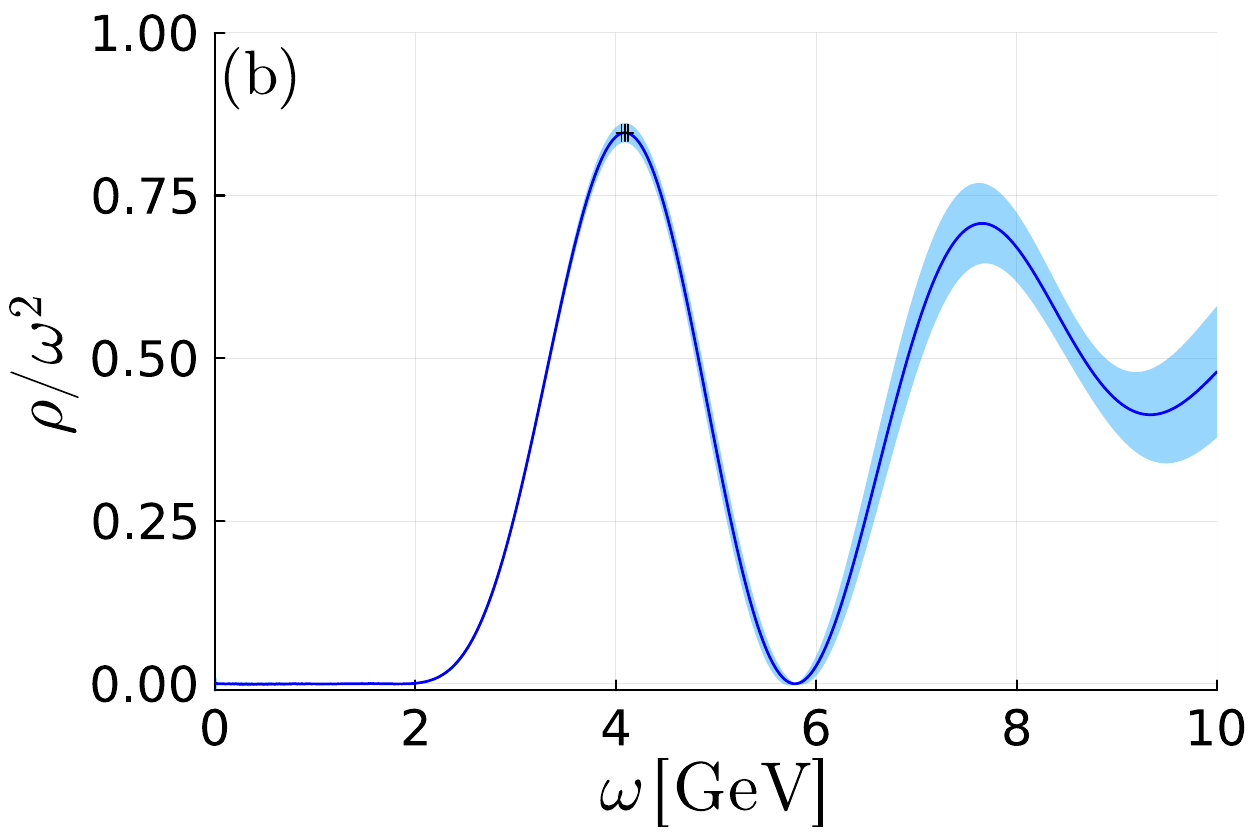}
  \end{minipage}
  \caption{
    Spectral functions in (a) vector and (b) pseudoscalar channels
    extracted from actual lattice QCD data for $T<T_{\mathrm{c}}$
    by using sparse modeling.
    The blue shaded areas represent the statistical errors of the spectral functions
    from Jackknife analyses,
    the blue solid lines represent the mean values,
    and the black horizontal error bars represent
    the uncertainty of the location of the first peak for each spectral function.
  }
  \label{fig:spf_LQCD_0.73Tc}
\end{figure}

\begin{figure}[tbp]
  \centering
  \begin{minipage}{.49\textwidth}
    \includegraphics[width=1.0\linewidth]{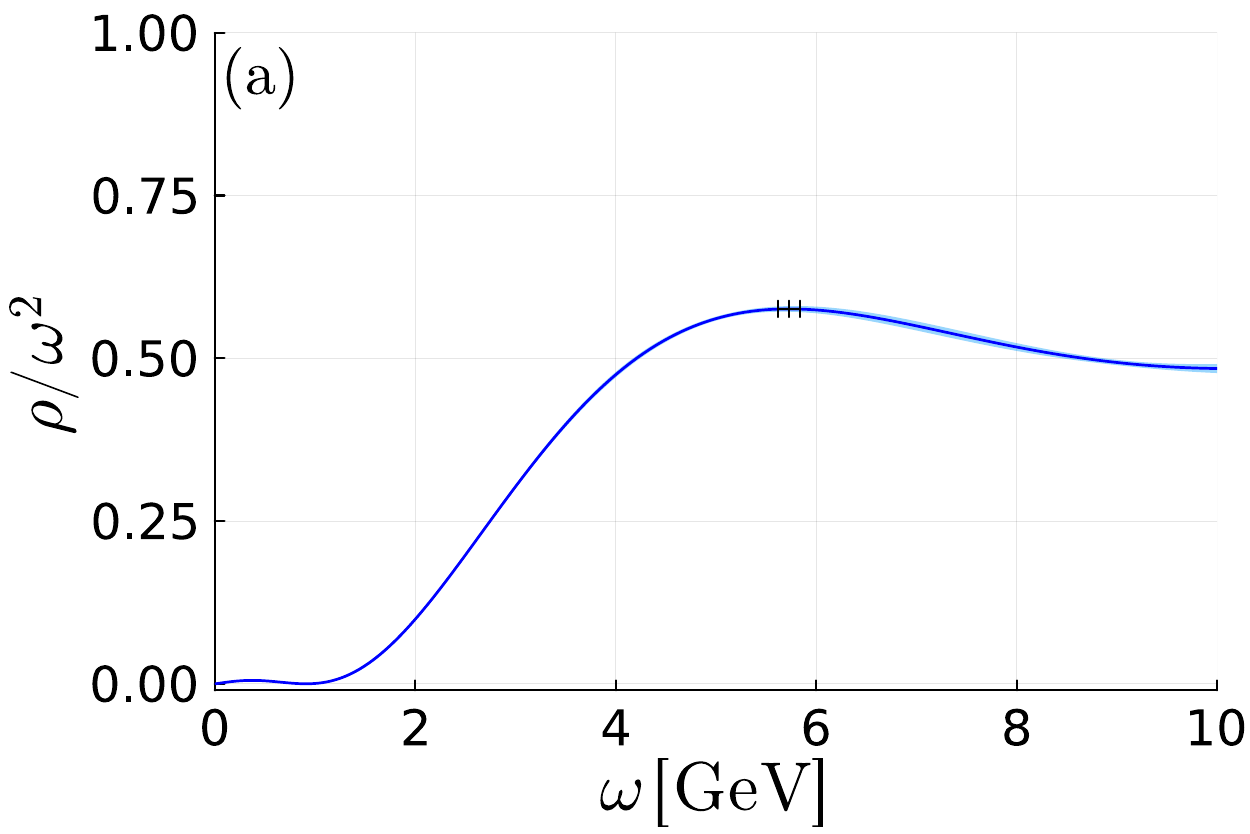}
  \end{minipage}
  \begin{minipage}{.49\textwidth}
    \includegraphics[width=1.0\linewidth]{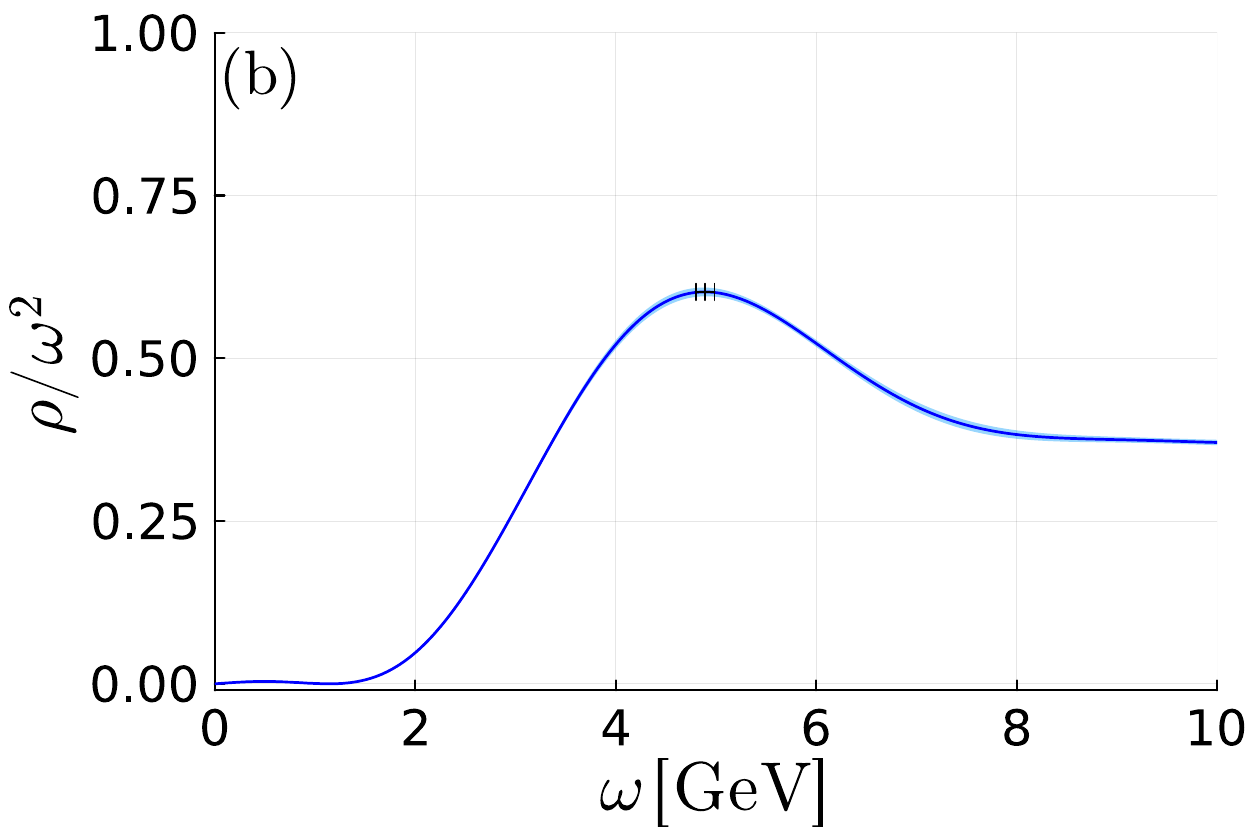}
  \end{minipage}
  \caption{
    Same as Fig.~\ref{fig:spf_LQCD_0.73Tc}
    but for $T>T_{\mathrm{c}}$.
    Figure (a) and (b) show the results in vector and pseudoscalar channels,
    respectively.
  }
  \label{fig:spf_LQCD_1.46Tc}
\end{figure}

\indent
Figure~\ref{fig:spf_LQCD_0.73Tc} shows
our results of the spectral functions
in (a) vector and (b) pseudoscalar channels for $T<T_{\mathrm{c}}$.
The blue shaded areas represent the statistical errors of the spectral functions
from Jackknife analyses,
the blue solid lines represent the mean values,
and the black horizontal error bars represent
the uncertainty of the location of the first peak for each spectral function.
The value of the spectral function increases around 2GeV.
The average value of the location of the first peak is 4.3 GeV in the vector channel
and 4.1 GeV in the pseudoscalar channel,
while those obtained from the maximum entropy method (MEM)
are about 3.48 GeV in the vector channel
and about 3.31 GeV in the pseudoscalar channel~\cite{Ding2012.PhysRevD.86.014509}.
Our result is a bit larger
compared to the result of the previous study.
\\
\indent
Figure~\ref{fig:spf_LQCD_1.46Tc} shows
the same as Fig.~\ref{fig:spf_LQCD_0.73Tc} but for $T>T_{\mathrm{c}}$.
Compared to the results for $T<T_{\mathrm{c}}$,
the peaks are broader and are located at higher energies.
The average values of the location of the first peak are 5.7 GeV in the vector channel
and 4.9 GeV in the pseudoscalar channel,
while those obtained from MEM
are about 4.7 GeV in the vector channel
and about 4.1 GeV in the pseudoscalar channel~\cite{Ding2012.PhysRevD.86.014509}.
Our results for higher temperatures are also a bit larger
compared to the results of the previous study.
In addition,
the transport peak does not appear.

\section{Summary}
We applied sparse modeling for extracting a spectral function
from a Euclidean-time meson correlation function.
Sparse modeling is a method that solves inverse problems
by considering only the sparseness of the solution we seek.
\\
\indent
First,
we tested sparse modeling with mock data
of the spectral function
which imitate possible charmonium spectral function
for $T<T_{\mathrm{c}}$ and $T>T_{\mathrm{c}}$
and checked applicability of sparse modeling.
This test confirmed
that increasing the number of data points of the correlation function
and reducing the noise level of errors of the correlation function
lead to output spectral functions closer to the input spectral function.
Despite the inclusion of transport peaks in the input spectral function
for $T>T_{\mathrm{c}}$,
our calculations do not yield the transport peak.
\\
\indent
Next,
we tried to extract the spectral functions from the charmonium correlation functions
in vector and pseudoscalar channels
for $T<T_{\mathrm{c}}$ and $T>T_{\mathrm{c}}$ obtained from lattice QCD.
Then, 
we got a spectral function with a broad peak around 4 GeV in each channel for $T<T_{\mathrm{c}}$,
which is a bit larger compared to the results in the previous study
using the maximum entropy method~\cite{Ding2012.PhysRevD.86.014509}.
For $T>T_{\mathrm{c}}$,
compared to the results for $T<T_{\mathrm{c}}$,
we got a spectral function with a broader peak around 5 GeV in each channel,
which is also a bit larger compared to the results from MEM.
In addition,
the transport peak does not appear.
In order to estimate the transport peak,
it may be necessary to make assumptions that extend beyond the sparse modeling,
including the shape of the transport peak.

\acknowledgments
We deeply grateful to H.-T. Ding for sharing lattice data.
The work of A.T. was partially supported by JSPS KAKENHI Grant Numbers 20K14479, 22H05111 and 22K03539.
A.T. and H.O. were partially supported by JSPS KAKENHI Grant Number 22H05112.
This work was partially supported by MEXT as ``Program for Promoting Researches on the Supercomputer Fugaku'' (Grant Number JPMXP1020230411, JPMXP1020230409).

\bibliographystyle{JHEP}
\bibliography{032_SpM}

\end{document}